\documentclass[12pt]{iopart}

\usepackage{graphicx}

\begin{document}

\title{Anisotropic covering of fractal sets}

\author{M. Wilkinson, H. R. Kennard}

\address{Department of Mathematics and Statistics, The Open University, Milton Keynes, MK7 6AA, England.}

\author{M. A. Morgan}

\address{Department of Physics, Seattle University, Seattle, WA98122, USA}

\begin{abstract}

We consider the optimal covering of fractal sets in a two-dimensional space using
ellipses which become increasingly anisotropic as their size is reduced.
If the semi-minor axis is $\epsilon$ and the semi-major axis is $\delta$, we set $\delta=\epsilon^\alpha$,
where $0<\alpha<1$ is an exponent characterising the anisotropy of the covers.
For point set fractals, in most cases we find that the number of points ${\cal N}$ which can be covered by an
ellipse centred on any given point has expectation
value $\langle {\cal N}\rangle \sim \epsilon^\beta$, where $\beta$
is a generalised dimension. We investigate the
function $\beta(\alpha)$ numerically for various sets, showing that
it may be different for sets which have the same fractal dimension.


\end{abstract}


\section {Introduction}
\label{sec: 1}

\begin{figure}
\centerline{\includegraphics[width=8.0cm]{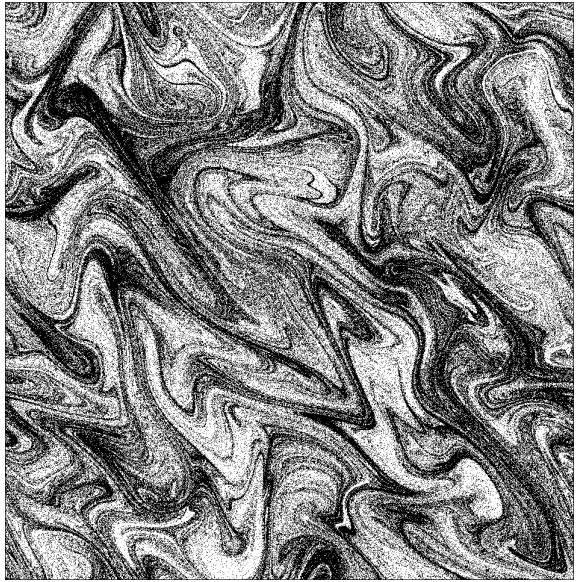}}
\caption{\label{fig: 1.1} Distribution of particles with significant inertia
moving in a two-dimensional area-preserving velocity field (the equations
of motion are specified in section \ref{sec: 3}). The particles tend to cluster
and for the parameters used in this simulation (see section \ref{sec: 3})
the correlation dimension is $D_2\approx 1.76$}
\end{figure}

\begin{figure}
\centerline{\includegraphics[width=8.0cm]{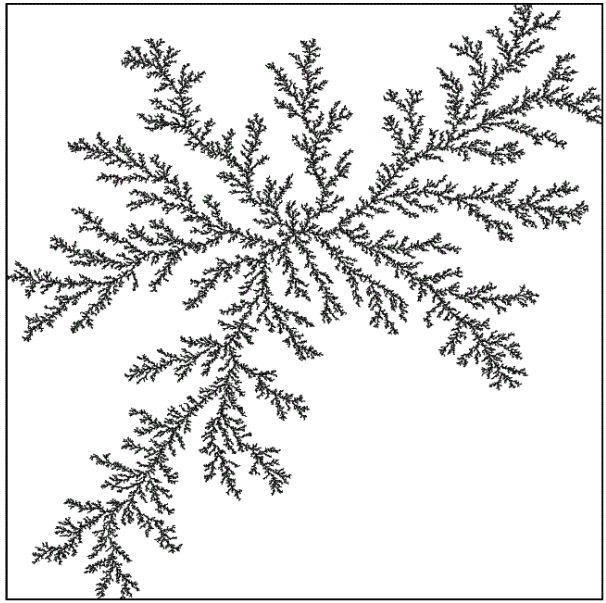}}
\caption{\label{fig: 1.2} Distribution of particles determined by a diffusion-limited
aggregation (DLA) process. This distribution has (approximately) the same fractal dimension as figure \ref{fig: 1.1}: $D_2\approx 1.71$}
\end{figure}

Figures \ref{fig: 1.1} and \ref{fig: 1.2} illustrate two different fractal point sets in two dimensions
(their origins will be described shortly). The fractal dimensions
(as defined in \cite{Man82,Fal90}) of these sets are very similar (the correlation
dimension are $D_2\approx 1.76$ and $D_2\approx 1.71$ respectively), but inspection of these figures
suggests that the fine-scale structure of these sets is quite different.
In this paper we consider whether it is possible to make a distinction between
the local structures of fractals by using covering sets which become
increasingly anisotropic as the covers are made smaller. We show that
this can distinguish different sets which have the same fractal
dimension. Moreover, we shall argue that the use of anisotropic covers can be
important in analysing the scattering of radiation from fractal distributions
of matter.

The sets illustrated in figures \ref{fig: 1.1} and \ref{fig: 1.2} are
point set fractals which arise as models for the distribution of particles
resulting from two distinct physical processes.
Figure \ref{fig: 1.1} illustrates a model of particles in a turbulent flow, in circumstances where
inertial effects are large enough to ensure that the particles are not simply
advected.  It is known that such systems may exhibit clustering \cite{Max87}
and that the attractor is a fractal measure \cite{Som+93,Wil+07}.
An enlargement of a subset of figure \ref{fig: 1.1} shows that the
particle distribution is {\sl locally} highly anisotropic, with the particles tending to be
clustered close to lines in the plane, even though the statistics of the distribution are
{\sl globally} rotationally invariant. This is consistent with the structure of strange
attractors in low-dimensional dissipative dynamical systems, where the attractor
has a local structure which is the Cartesian product of a line and a one-dimensional
Cantor set \cite{Ott02}. The second example, figure \ref{fig: 1.2}, is a  diffusion-limited
aggregation (DLA) cluster in two dimensions \cite{Wit+81}. The local structure
of this set cannot be approximated as a Cartesian product. Also, if a subset
of a DLA cluster such as figure \ref{fig: 1.2} is examined at a higher magnification,
it does not appear to be more markedly anisotropic
in structure. Thus it appears as if some fractal sets, such as that illustrated
in figure \ref{fig: 1.1}, may exhibit a pronounced anisotropy under magnification, whereas
this effect is much less pronounced (or possibly absent) in others, such as figure \ref{fig: 1.2}.

Throughout this paper we confine our attention to point sets in the plane, but
generalisations to higher dimensions are obvious. We use the following approach to
characterise the local structure of a point set. Take a given element of the set, and consider an
ellipse centred on this point, with its semi-minor axis of length $\epsilon$ and
its semi-major axis of length $\delta=\epsilon^\alpha$, where $0<\alpha<1$.  We assume
$\delta\ll \xi$ and $\epsilon \ll \xi$, where $\xi$ is the characteristic
lengthscale of the system below which fractal
scaling is observable, and that $\epsilon \gg \epsilon_0$,
where $\epsilon_0$ is the lengthscale
where the fractal scaling is cut off by the finite number of points
sampling the fractal measure. We then choose the
orientation of the cover so that it maximises the
number of other points which are
contained in this ellipse. We denote the number of points
under this optimally-oriented cover by ${\cal N}$.
We repeat this for ellipses centred on other randomly selected
points in the set, and compute the average value $\langle {\cal N}\rangle$ of the number of points
which can be covered. In most of the examples of point-set fractals
which we investigated, the mean number of points in this ellipse is found
to have a power-law dependence:
\begin{equation}
\label{eq: 1.1}
\langle {\cal N}(\epsilon,\alpha)\rangle \sim \epsilon^{\beta(\alpha)}\ ,\ \ \ \ \delta=\epsilon^\alpha
\end{equation}
where the exponent $\beta$ depends upon $\alpha$. In the case where $\alpha=1$, the ellipse
is a circle, so that this case reduces to a definition of the correlation dimension: $D_2=\beta(1)$
(compare with the discussion of the correlation dimension in \cite{Gra+84}).
In general $\beta$ must decrease monotonically as $\alpha$ decreases.

A related approach to the characterisation of fractal sets was proposed
by Grassberger and co-workers \cite{Gra+84,Gra85,Pol+88,Gra+88},
who considered covering a set
(which is embedded in a $d$-dimensional space) with
$d$-dimensional ellipsoids, with principal axes $\epsilon_i$, ordered so that
$\epsilon_1>\epsilon_2>\ldots >\epsilon_d$. Their work emphasises the
case where the local structure of the fractal is a Cartesian product of sets,
with dimensions $\Delta_i$, ordered so that $\Delta_1>\Delta_2>\ldots>\Delta_d$. It is asserted that
the ellipsoids cover most efficiently when they align with the principal axes, such that
the longest axes align with the directions of the highest density sets. According to this hypothesis
the number of points covered is expected to satisfy
\begin{equation}
\label{eq: 1.2}
\langle {\cal N}\rangle \sim \epsilon_1^{\Delta_1}\epsilon_2^{\Delta_2}\epsilon_3^{\Delta 3}\ldots \epsilon_d^{\Delta_d}
\ .
\end{equation}
Examining the dependence of ${\cal N}$ upon the $\epsilon_i$ would allow the \emph{partial dimensions} $\Delta_i$
to be determined.
This approach was mentioned in several papers \cite{Gra+84,Gra85,Pol+88,Gra+88}, with the motivation
to characterise a fractal set by means of its partial dimensions, $\Delta_i$, satisfying
$\sum_{i=1}^d \Delta_i=D_2$. These works do not prescribe how (or whether) the ratio
$\epsilon_{i+1}/\epsilon_i$ approaches zero as $\epsilon_1\to 0$. In our work
this limiting behaviour is specified by the parameter $\alpha$. Our numerical
investigations encompass fractal sets which have a Cartesian product structure, and
those which do not.

\begin{figure}
\centerline{\includegraphics[width=5.0cm]{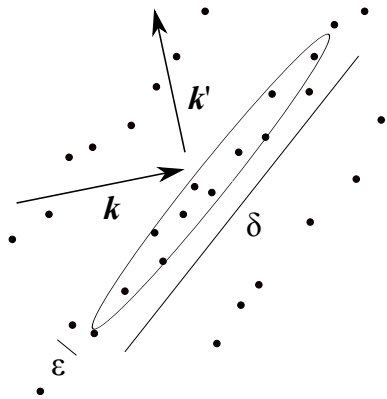}}
\caption{\label{fig: 1.3} Scattering of a beam of light with wavevector 
$\mbox{\boldmath$k$}$, width $\delta$ and wavelength $\epsilon$:
a cluster of ${\cal N}$ particles covered by an $\epsilon$-$\delta$ ellipse scatters light
coherently when the condition for specular reflection is satisfied. The wavevector
$\mbox{\boldmath$k$}'$ satisfies the condition for specular reflection if the
major axis of the ellipse is perpendicular to $\mbox{\boldmath$k$}-\mbox{\boldmath$k$}'$.
In this case the scattered intensity from the particles under the cover is increased by a
factor of order ${\cal N}$. }
\end{figure}

In the case of point set fractals which represent a physical distribution of particles, the existence
of a highly anisotropic local structure has important physical implications for the scattering of light.
It is known that s-wave scattering of light from a fractal point set gives rise to an
algebraic relation between the scattering wavenumber $k$ and the scattered intensity $I$, such that
\begin{equation}
\label{eq: 1.3}
I(k)\sim k^{-D_2}
\ .
\end{equation}
Heuristic arguments supporting this are developed in \cite{Sin89}; see \cite{Det+93}
for a discussion of some of the complications which can arise in verifying this suggestion.
The distribution of this scattering intensity may, however, be highly inhomogeneous.
In particular, if the particles have a strong tendency to accumulate along
lines in two dimensions (like the example shown in figure \ref{fig: 1.1}), or
on planes in higher dimensions, light may scatter
specularly from these structures. We argue that this motivates the investigation
of the anisotropic covering with ellipsoids. Consider weak scattering of light with wavelength
$\epsilon$ which propagates as a beam of width $\delta$. When the path length
for light scattered from different particles is large compared to $\epsilon$, the scattering of light from
${\cal N}$ particles is incoherent, so that the contribution to the scattered
intensity is $I\sim {\cal N}$. If, however, an ellipsoid of major axis $\delta$
and minor axis $\epsilon$ can be aligned to cover ${\cal N}$ particles, then there will
exist directions where the path length difference is less than one wavelength, so that this
set of ${\cal N}$ particles scatters light coherently (see figure \ref{fig: 1.3}).
 In these directions where the condition for
specular reflection is satisfied by the optimal covering ellipsoid, there is a greatly increased intensity
$I\sim {\cal N}^2$.

We have argued that light scattering from fractal sets may be
extremely inhomogeneous because of specular scattering from
particles which accumulate on surfaces. This motivated us to
study the covering of a fractal by anisotropic covering sets, and
the results are reported here. We investigate the dependence of the
generalised dimension $\beta$ upon the anisotropy exponent $\alpha$.
We show that the form of the function $\beta(\alpha)$ can distinguish between different
fractal sets which have the same value of the correlation dimension $D_2=\beta(1)$.
The description of light scattering
proved to be a very complex issue, which will be considered elsewhere.

\section{Some elementary estimates}
\label{sec: 2}

Before discussing our numerical investigations, we consider some simple arguments
about the form of the generalised dimension $\beta(\alpha)$, defined by equation (\ref{eq: 1.1}).

\begin{figure}
\centerline{\includegraphics[width=6.5cm]{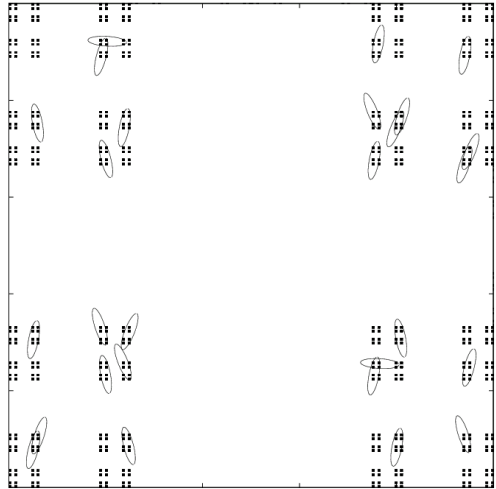}}
\caption{\label{fig: 2.1} In the case where the fractal set is a Cartesian product of
two Cantor sets, with dimensions $D_x$ and $D_y$ in the $x$ and $y$ directions, the optimal
covering ellipse might be expected to have its major axis aligned with the direction corresponding
to the denser set. In the case $D_y>D_x$, the major axis of the ellipse would be expected to align with
the $y$-axis (vertical in this figure). Here we illustrate a sample of the actual optimal covers, which have a distribution of angles of their
principal axes.}
\end{figure}

First we address the issue of whether the exponent $\beta$ exists.  When $\alpha=1$, the exponent
$\beta$ coincides with the correlation dimension of the set. For other values of
$\alpha$, we do not know of any general argument which proves that the dependence of the
optimal covering ${\cal N}$ has a power law relation to the size $\epsilon$ of the covering
elements. For most of the point-set fractals which we examined we did find good
numerical evidence that $\langle {\cal N}\rangle \sim \epsilon^\beta$ for small values of $\epsilon$
(extending down towards values of $\epsilon$ where the discrete sampling of the set becomes a limitation).
The exceptions occurred for some cases of the motion of inertial particles in a random flow, and for
some of the sets considered in section \ref{sec: 4}, including the example illustrated in
figure \ref{fig: 2.1}: these are discussed in sections \ref{sec: 3} and \ref{sec: 4} respectively.

Next we give an upper bound on $\beta(\alpha)$. Let us consider the expectation
$\langle {\cal N}\rangle$ for the case where ${\cal N}$ is independent of the orientation
of the ellipse. The total number of particles in a disc of radius $\delta$ is proportional to
$\delta ^{D_2}$. If the set is locally isotropic then the coverage is independent
of the angle of the major axis of the ellipse, and a fraction $\epsilon/\delta$ of these
points lie in the ellipse. Recalling that $\delta=\epsilon^\alpha$, for this isotropic fractal we have
$\langle {\cal N}\rangle \sim \delta^{D_2-1}\epsilon \sim \epsilon^{1-\alpha+\alpha D_2}\equiv \epsilon^{\beta_+(\alpha)}$, so that
\begin{equation}
\label{eq: 2.1}
\beta_+(\alpha)=1+(D_2-1)\alpha
\ .
\end{equation}
is an upper bound on the exponent $\beta$.

We were not able to obtain a precise and non-trivial lower bound for $\beta(\alpha)$, but the following
argument suggests a possible form for a lower bound.
Consider the case where the point fractal samples a Cartesian
product of two one-dimensional Cantor sets, with dimensions $D_x$ and $D_y$. In the following
we assume that $D_y\ge D_x$. Because the set is \lq denser' in the direction of the $y$-axis,
we expect that the optimal alignment of each ellipse is when its major
axis is aligned with the $y$-axis (the same hypothesis was proposed in \cite{Gra+84,Gra85}).
The expected number of particles captured by a
covering ellipse is then $\langle {\cal N}\rangle \sim \delta^{D_y}\epsilon^{D_x}=\epsilon^{\alpha D_y+D_x}$, so that the dimension of this product set is
\begin{equation}
\label{eq: 2.2}
\beta(\alpha)=\alpha D_y+D_x
\ .
\end{equation}
Now consider the smallest possible dimension which could be achieved
according to this argument, if we allow the dimensions $D_x$, $D_y$ of the component sets to vary so that
$D_x+D_y=D_2$. The greatest number of particles in the ellipse, and hence the smallest dimension,
is obtained by setting $D_y=1$ and $D_x=D_2-1$. This gives a putative lower bound to the dimension:
\begin{equation}
\label{eq: 2.3}
\beta_-(\alpha)=D_2-1+\alpha
\ .
\end{equation}
The assumption that the covering ellipses align precisely with the $y$-axis is not
really correct, as evidenced by figure \ref{fig: 2.1}. It is plausible that the
probability for an ellipses to be significantly mis-aligned decreases
as $\epsilon \to 0$, but we were not able to obtain conclusive numerical
evidence. We note that the argument leading to the proposed lower
bound, $\beta_-(\alpha)$, is very similar to that presented in \cite{Gra+84,Gra85}
to motivate the concept of partial dimensions.

\begin{figure}
\centerline{\includegraphics[width=4.5cm]{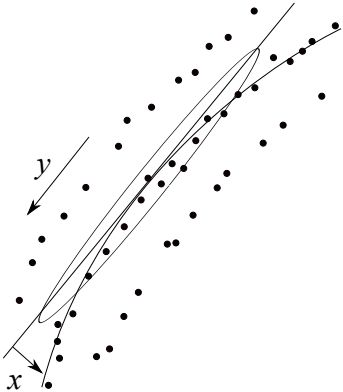}}
\caption{\label{fig: 2.2} In many dynamical systems the attractor is locally a Cartesian product of a Cantor set
and a line. If this line is curved, an ellipse may cease to be an efficient cover as
$\alpha$ is decreased. The $x$- and $y$-coordinates are referred to in section \ref{sec: 2}.}
\end{figure}

In addition to the fact that the optimal covering ellipses do not align
perfectly with the preferred axes, there is a further complication which could affect the argument
leading to the estimate in equation (\ref{eq: 2.3}). In sets such as that
illustrated in figure \ref{fig: 1.1}, the local structure is a Cartesian product of a
line and a one-dimensional Cantor set. The line is, however, curved.
It is interesting to consider whether this curvature can alter the estimate
in equation (\ref{eq: 2.3}). In order to consider the
effect of this curvature, introduce two coordinates: $y$ is a coordinate for the
expanding (unstable) manifold centred on the reference particle and $x$ is a coordinate
for the stable manifold. Because the manifolds are curved, the equation of the unstable
manifold will be of the form $x\sim Cy^2$ for small $y$, where $C$ is a constant. Consider
a family of ellipses, with fixed $\delta$ and decreasing $\alpha$ which are aligned
so as to provide the optimal cover for a cluster centred on a reference particle.
As $\alpha$ is reduced, the curvature of the unstable manifold can become important, because
it will take a cluster of particles outside the covering ellipse (see figure \ref{fig: 2.2}). This happens when
$x(\delta)\sim \epsilon$, that is when $C\epsilon^{2\alpha}\sim \epsilon$.
Accordingly, we might propose that the optimal
covering strategy of aligning the principal axis of the ellipses with the unstable
manifold starts to break down when at a critical value of the exponent, $\alpha_{\rm c}=\frac{1}{2}$.
On the basis of this argument we would expect that $\beta(\alpha)$ might exceed equation (\ref{eq: 2.3})
when $\alpha<\frac{1}{2}$, in cases where the fractal is generated by a dynamical system for which
the lines representing the unstable manifold are curved.

\section{Numerical investigations of dynamical fractals}
\label{sec: 3}

Figure \ref{fig: 1.1} illustrates the distribution of independently moving inertial particles in a random velocity
field. The equations of motion for the position of a given particle are \cite{Max87}
\begin{equation}
\label{eq: 3.1}
\dot{\mbox{\boldmath$r$}}=\mbox{\boldmath$v$}
\ ,\ \ \
\dot{\mbox{\boldmath$v$}}=-\gamma[\mbox{\boldmath$v$}-\mbox{\boldmath$u$}(\mbox{\boldmath$r$}(t),t)]
\end{equation}
where $\gamma$ is the rate at which the particles relax towards
the fluid velocity, and where $\mbox{\boldmath$u$}(\mbox{\boldmath$r$},t)$ is a randomly fluctuating
velocity field satisfying the incompressibility condition
$\mbox{\boldmath$\nabla$}\cdot \mbox{\boldmath$u$}=0$. Particles in the fluid
flow cluster if the damping timescale $\gamma^{-1}$ is comparable to a timescale
characterising the velocity field. In the simulations used in this paper, we used
a random vector field with a very small correlation time, using the same definitions as in
\cite{Wil+10}, where the importance of inertial effects is characterised
by a dimensionless parameter, which was referred to as $\varepsilon$ in that work, but
which is denoted by $\eta$ in this paper. It is defined in terms of the correlation
function of the velocity gradient experienced by a particle with trajectory $\mbox{\boldmath$r$}(t)$:
\begin{equation}
\label{eq: 3.2}
\eta^2=\frac{1}{2\gamma}\int_{-\infty}^\infty {\rm d}t\ \left\langle \left(\frac{\partial u_x}{\partial x}\right)(\mbox{\boldmath$r$}(t),t)\left(\frac{\partial u_x}{\partial x}\right)({\bf 0},0)\right\rangle
\ .
\end{equation}
 In figure \ref{fig: 1.1}, we showed
a realisation of the long-time dynamics. The velocity is periodic on the
unit square, and was generated from a random stream function $\psi(\mbox{\boldmath$r$},t)$. The statistics of the stream function are $\langle \psi(\mbox{\boldmath$r$},t)\rangle=0$,
$\langle \psi(\mbox{\boldmath$r$},t)\psi({\bf 0},0)\rangle=A^2\exp(-|\mbox{\boldmath$r$}|^2/2\xi^2)\exp(-|t|/\tau)$, with
$\xi=0.1$, $\tau$ small and $A$ chosen such that $\eta=0.1$. The correlation
dimension for this value of $\eta$ is $D_2\approx 1.76$ \cite{Wil+10}.

\begin{figure}
\centerline{\includegraphics[width=10.0cm]{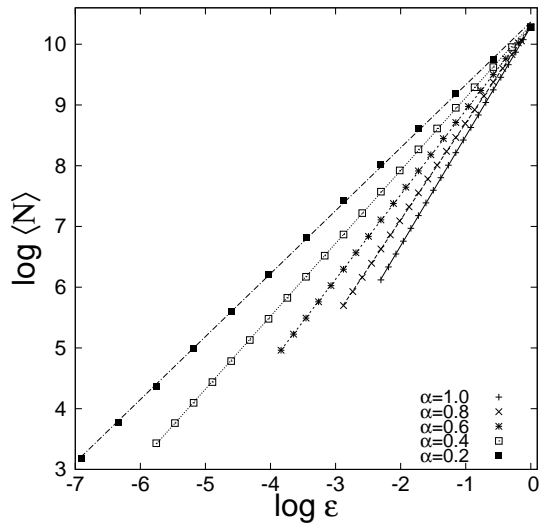}}
\caption{\label{fig: 3.1} The mean number of points in an optimal
cover, $\langle {\cal N}\rangle$, as a function of $\epsilon$, for various
values of $\alpha$. The slope increases monotonically with $\alpha$,
which ranges from $\alpha =0.2$ to $\alpha=1$ in increments of $0.2$.
These data, for the random flow model with $\eta=0.1$,
show excellent fits to a power law over a wide range of $\epsilon$.}
\end{figure}

We examined whether the mean value of the optimal
covering number, $\langle {\cal N}\rangle$, shows a power-law dependence upon
$\epsilon$. The data for $\eta=0.1$, shown in figure \ref{fig: 3.1}, are a good fit to a power-law
for values of $\alpha$ as low as $0.2$. At very small values of $\alpha$, the area
of the ellipses decreases very rapidly as $\epsilon \to 0$, so that the values
of $\langle {\cal N}\rangle$ become too small to give reliable results.

\begin{figure}
\centerline{\includegraphics[width=10.0cm]{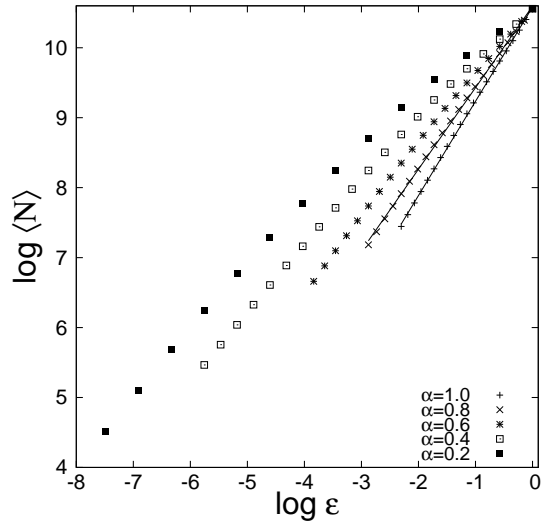}}
\caption{\label{fig: 3.2} Same as figure \ref{fig: 3.1}, except $\eta=0.4$,
where $D_2\approx 1.36$. In this case the covering data are not a good fit to
a power-law in $\epsilon$ for $\alpha \le 0.75$.}
\end{figure}

For the inertial particles model we found a number of cases where the covering data was
not well fitted by a power-law in $\epsilon$.  This occurs for small values of $\alpha$ and for values of $\eta$ where the value of $D_2$
is close to its minimum, which is $D_2\approx 1.35$ at $\eta\approx 0.35$. Figure
\ref{fig: 3.2} illustrates the case where the fit to a power-law was the least good.

\begin{figure}
\centerline{\includegraphics[width=12.0cm]{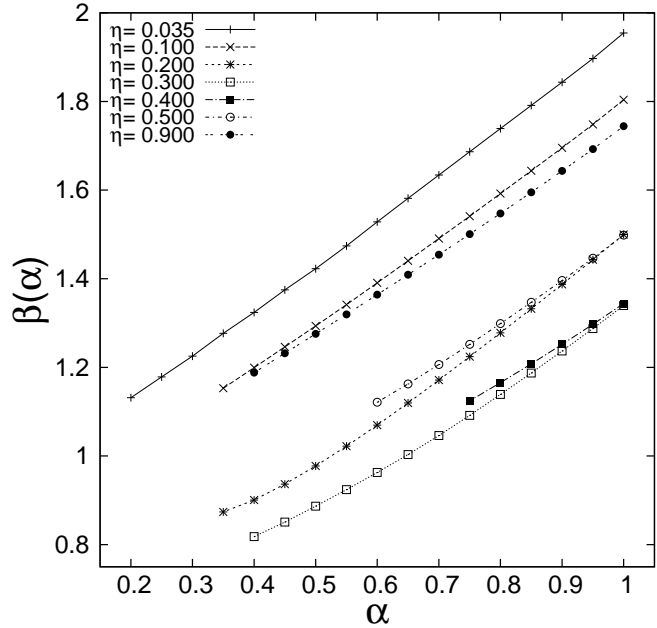}}
\caption{\label{fig: 3.3} $\beta(\alpha)$ for inertial particles in a random
velocity field, using a model defined in \cite{Wil+10}. The curves are labelled
according to the parameter $\eta$ which quantifies the importance of
particle inertia.}
\end{figure}

In figure \ref{fig: 3.3} we exhibit the $\beta(\alpha)$ curves
for this systems at several different values of the inertial
parameter $\eta$.
Fractal attractors of dynamical systems typically have a local structure which is a
Cartesian product of a Cantor set and a line. Following the discussion in
section \ref{sec: 2} we therefore expect that
the exponent $\beta(\alpha)$ should be given by equation (\ref{eq: 2.3}),
that is $\beta(\alpha)\approx \beta_-(\alpha)$.
We find, however, that $\beta_-(\alpha)$ is not a very good approximation, and
figure \ref{fig: 3.3} shows that different $\beta(\alpha)$ curves may be obtained
for cases with the same fractal dimension (these arise because $D_2$  has a minimum
with respect to varying $\eta$) .
In section \ref{sec: 2} we also suggested that the generalised
dimension $\beta(\alpha)$ might be higher than $\beta_-(\alpha)$ when
$\alpha \le \frac{1}{2}$ because the unstable manifold is curved. Figure \ref{fig: 3.3}
does not show evidence of any discontinuous change at $\alpha=\frac{1}{2}$.

\begin{figure}
\centerline{\includegraphics[width=12.0cm]{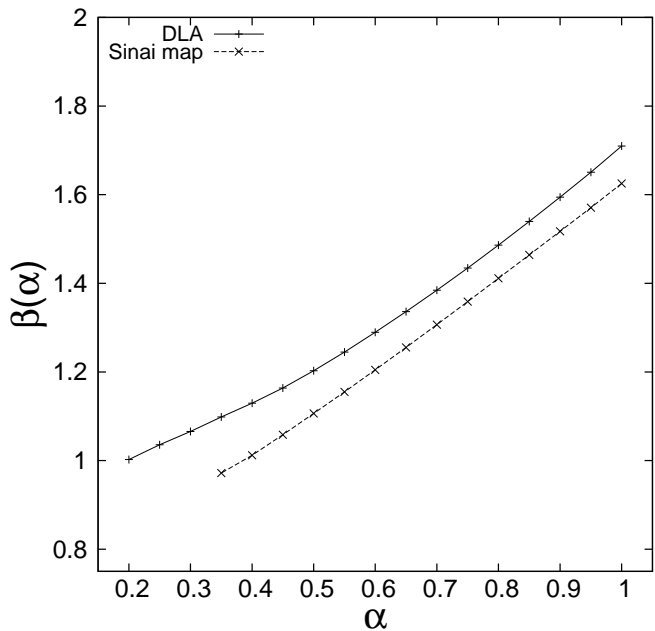}}
\caption{\label{fig: 3.4} $\beta(\alpha)$ for a diffusion limited aggregation (DLA) cluster
with $2.7\times 10^5$ particles, and for the Sinai map with $a=0.35$.}
\end{figure}

We also investigated $\beta(\alpha)$ for fractals generated by
two other dynamical processes. We simulated diffusion limited
aggregation (DLA, one realisation of which was illustrated in figure \ref{fig: 1.2}),
 in the usual manner \cite{Wit+81}:  points make a random walk on a lattice
until they contact the cluster,
after which they are frozen and become part of the growing set. In the case
of isotropic diffusion, the correlation dimension of the resulting cluster
is $D_2\approx 1.71$. The values of the slopes $\beta$ extracted from
least-squares fits similar to those in figure \ref{fig: 3.1} are plotted in
figure \ref{fig: 3.4} as a function of $\alpha$.

\newpage
We also investigated the fractal generated by the Sinai map, defined by
\begin{eqnarray}
\label{eq: 3.3}
x_{n+1}&=&x_n + y_n + a \cos(2\pi y_n){\rm  mod}\, 1
\nonumber \\
y_{n+1}&=&x_n + 2 y_n{\rm  mod}\, 1
\end{eqnarray}
with $a=0.35$, which has an attractor with correlation dimension $D_2\approx 1.62$.
The $\beta(\alpha)$ curve for this map is also shown in figure \ref{fig: 3.4}.

\section{Sierpinski substitution fractals}
\label{sec: 4}

The examples of dynamical fractals
which we considered in section \ref{sec: 3} are all multifractal sets,
and it is desirable to investigate $\beta(\alpha)$ for a model
which is a simple fractal, avoiding the complications that arise
when dealing with multifractal sets \cite{Hal+86}. Here we construct a
class of generalisations of the Sierpinski carpet, which are simple
fractals rather than multifractal measures. The construction that we use
is closely related to one proposed independently by Bedford
\cite{Bed84} and McMullen \cite{McM84}. We show that different
elements from this class of sets can have different $\beta(\alpha)$
functions despite having precisely the same value of $D_2=\beta(1)$.

We generate an approximation to a fractal set by a hierarchical
process consisting of $n$ generations. We generate a set of $M^n$ points,
where $M$ is an integer, as follows.
The points $\mbox{\boldmath$x$}_k$ lie in the unit square $[0,1]\otimes [0,1]$, and have
coordinates of the form $(x_i,y_i)=(i/N_1^n,j/N_2^n)$, where $N_1$, $N_2$ are positive integers satisfying
$N_1 N_2>M$.

We define a \lq masking matrix' ${\bf F}$ with elements $F_{ij}$ as an
$N_1\times N_2$ matrix which has elements which are either $1$ or $0$,
and we let $M$ be the number of non-zero elements of ${\bf F}$.
We construct the model set by the following recursive construction.
Consider the \lq first generation' set of $M$ points
$\mbox{\boldmath$x$}_k$, labelled by an index $=1\ldots M$,
where a point is placed at $((i-1)/N_1,(j-1)/N_2)$ if
$F_{ij}=1$. At the next generation \emph{each} of these points is replaced by a
set of $M$ points, based on a lattice with spacings $N_1^{-2}$ and $N_2^{-2}$
in the $x$ and $y$ directions respectively, which are selected by the criterion
that $F_{ij}=1$. In general,
after $n$ generations each point $\mbox{\boldmath$x$}_k$ is replaced by
$M$ points with positions $\mbox{\boldmath$x$}_{k'}$, where $k'$ is an index
of the $(N_1N_2)^{n+1}$ points, with positions
\begin{equation}
\label{eq: 4.1}
\mbox{\boldmath$x$}'_{k'}=\mbox{\boldmath$x$}_k+\left(\frac{i-1}{N_1^{n+1}},\frac{j-1}{N_2^{n+1}}\right)
\ .
\end{equation}
A point labelled by $(i,j)$ added to the set if and only if $F_{ij}=1$.

\begin{figure}
\centerline{\includegraphics[width=9.0cm]{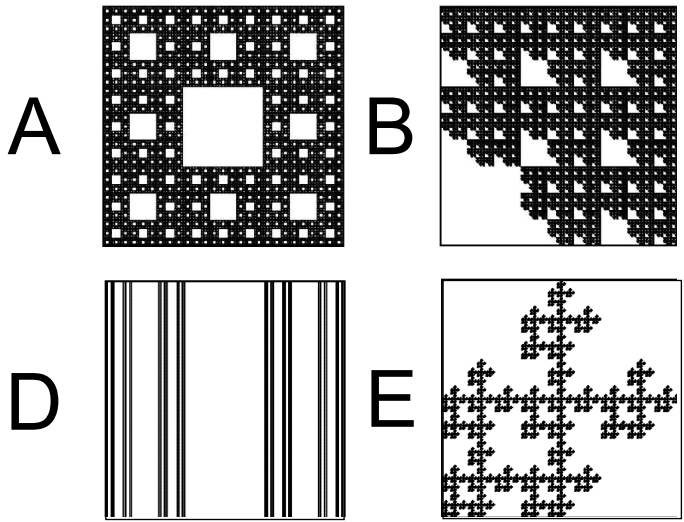}}
\caption{\label{fig: 4.1} Examples of fractal sets generated
by the construction defined in section \ref{sec: 4}. In all of these
examples, $N_1=N_2=3$. The Cantor set is then defined by listing the
zero elements of the masking matrix (the labels correspond to the cases
considered in figure \ref{fig: 4.2}):
(A) $F_{22}=0$, $D={\rm ln}\,8/{\rm ln}\,3$,
(B) $F_{11}=0$, $D={\rm ln}\,8/{\rm ln}\,3$,
(D) $F_{21}=F_{22}=F_{23}=0$, $D={\rm ln}\,6/{\rm ln}\,3$,
(E) $F_{13}=F_{31}=F_{33}=0$, $D={\rm ln}\,6/{\rm ln}\,3$
.}
\end{figure}

As an example, consider the case where $N_1=N_2=3$ and where $F_{22}=0$ is the
only element of ${\bf F}$ which is equal to zero, so that $M=8$. Iterating this construction gives
a version of the Sirepinski carpet set, illustrated in figure \ref{fig: 4.1}A, with dimension
$D={\rm ln}\,8/{\rm ln}\,3$. By varying
the zero elements of the masking matrix, we can generate many other Cantor
sets, some examples of which are illustrated in the other panels of figure \ref{fig: 4.1}.
The resulting sets are clearly simple fractals (as opposed to multifractals). By making other choices of the masking
matrix we can construct other Cantor sets with dimension
\begin{equation}
\label{eq: 4.2}
D=\frac{2\,{\rm ln}\,M}{{\rm ln}\,(N_1\,N_2)}
\ .
\end{equation}
This construction allows us to create distinct fractal sets with exactly the same dimension,
such as in panels A and B or panels D and E of figure \ref{fig: 4.1}. Moreover, by a suitable choice of the
masking matrix, we can generate fractal sets which are Cartesian products (such as figure \ref{fig: 4.1}D), as well as those
which are not (such as figures \ref{fig: 4.1}A, B, E).

\begin{figure}
\centerline{\includegraphics[width=15.0cm]{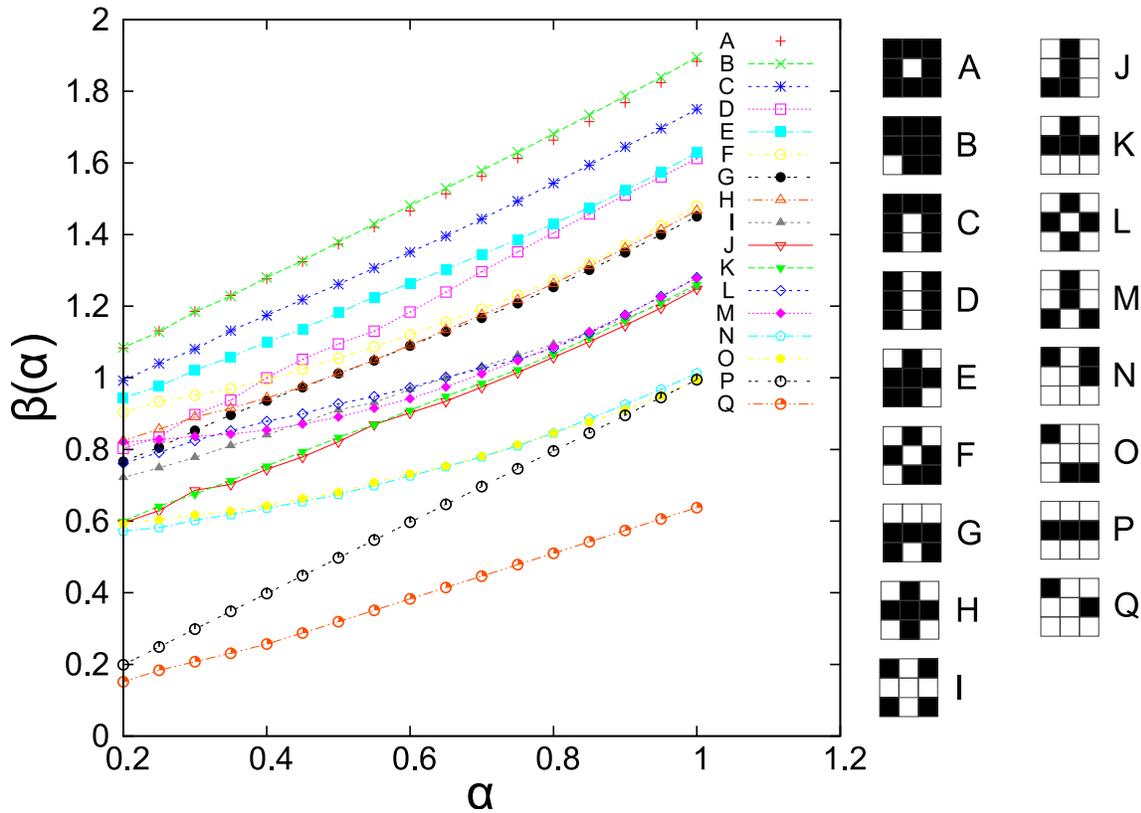}}
\caption{\label{fig: 4.2} $\beta(\alpha)$ for the generalised Sierpinski model.
The curves are labelled by a key giving the zero elements of the masking matrix
in white.}
\end{figure}

We investigated the function $\beta(\alpha)$ for sets which are produced by
this generalised Sierpinski construction. The results are illustrated in figure \ref{fig: 4.2},
for seventeen sets produced using a $3\times 3$ masking matrix. The key at the right hand side
of the figure indicates the pattern of deletions in the masking matrix, ordered by the number
of deleted points.

First we discuss those sets which are a
Cartesian product. These include examples  D,  I, L, P and Q in figure \ref{fig: 4.2}.
The simplest example is set D in figures \ref{fig: 4.1} and \ref{fig: 4.2}.
For this set, equation (\ref{eq: 2.3}) predicts that $\beta(\alpha)=D-1+\alpha$, which shows excellent
agreement with figure \ref{fig: 4.2}.
By setting $N_1=N_2=3$ and $F_{12}=F_{21}=F_{23}=F_{32}=F_{22}=0$ we produce a set with dimension
$D={\rm ln}\,4/{\rm ln}\,3=2\,{\rm ln}\,2/{\rm ln}\,3$, which is a Cartesian product of two Cantor
sets of dimension $D_x=D_y=D/2={\rm ln}\,2/{\rm ln}\,3$. This is example I in figure \ref{fig: 4.2}.
Example  L is closely related: this set is similar to example  I, rotated
by $\pi/4$. These data show quite poor agreement with the prediction from equation (\ref{eq: 2.2}), from which we expect $\beta(\alpha)=D(1+\alpha)/2$ (but good agreement with each other).
The other two examples in figure \ref{fig: 4.2} which are Cartesian products are very simple:  P
is a Cartesian product of a line and a point, and  Q is the product of a Cantor set of dimension ${\rm ln}\,2/{\rm ln}\,3$
and a point. For these examples there is excellent agreement with the predictions of equations (\ref{eq: 2.3}) and (\ref{eq: 2.2}),
which indicate straight lines of slope unity and ${\rm ln}\,2/{\rm ln}\,3$ respectively.

Figure \ref{fig: 4.2} also shows $\beta(\alpha)$ in cases where the set is not a Cartesian product.
These differ from the data for the Cartesian product sets. They can be organised
into sets which have apparently identical $\beta (\alpha)$ curves. In most of the cases
examined in figure \ref{fig: 4.2}, sets which have the same value of $M$ (and hence of $D_2$)
have $\beta(\alpha)$ curves which are identical, to within numerical fluctuations. Examples
of such groups are $(M=8:  {\rm A,B})$, $(M=5: {\rm G}, {\rm F})$, $(M=4: {\rm J}, {\rm K})$, $(M=3: {\rm N}, {\rm  O})$.
Note, however, that for $M=4$, set M is not a Cartesian product and yet has a $\beta(\alpha)$ curve which is clearly different from sets J and  K.

\begin{figure}
\centerline{\includegraphics[width=10.0cm]{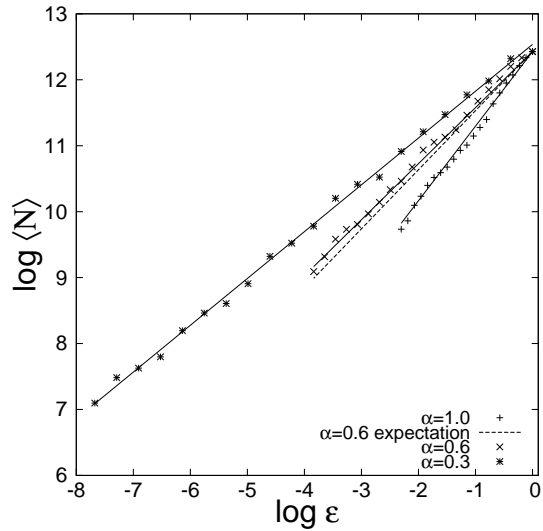}}
\caption{\label{fig: 4.3} For the set illustrated in figure \ref{fig: 2.1}, it is
difficult to fit the exponent $\beta(\alpha)$ because of an oscillation in the
dependence of ${\rm ln}\,\langle {\cal N}\rangle$ upon ${\rm ln}\,\epsilon$.}
\end{figure}

\begin{figure}
\centerline{\includegraphics[width=12.0cm]{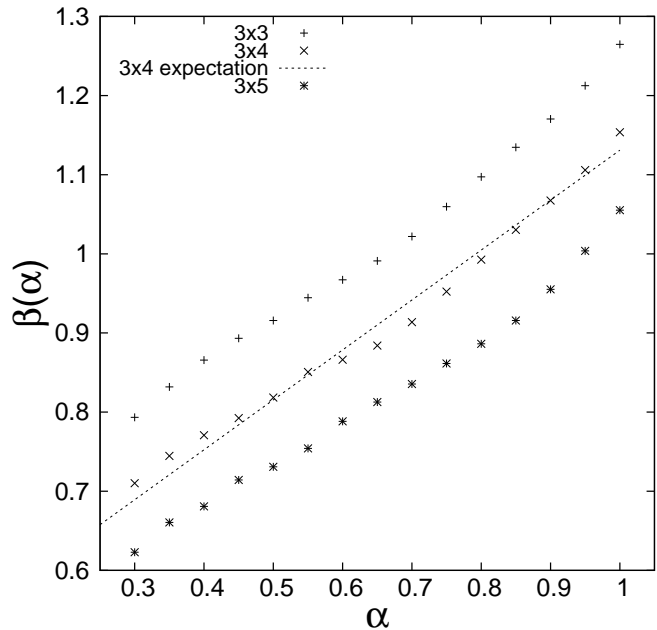}}
\caption{\label{fig: 4.4} $\beta(\alpha)$ for three cases of the generalised
Sierpinski model which are Cartesian products of Cantor sets, based upon
$3\times 3$, $4\times 3$ and $5\times 3$ deletion matrices.}
\end{figure}

We noted that examples I and  L in figure \ref{fig: 4.2}, which are Cartesian products of two one-dimensional
Cantor sets, had $\beta(\alpha)$ functions which show quite poor agreement with the expected
result, equation (\ref{eq: 2.2}).  It appears possible that this anomaly might arise because
these sets are a degenerate case, where $D_x=D_y$. We therefore also considered two
examples which are a Cartesian product of two Cantor sets with different dimensions, namely
$N_1=4$, $N_2=3$, with non-zero elements $F_{11}=F_{13}=F_{41}=F_{43}=1$
(which is the set illustrated in figure \ref{fig: 2.1}), and $N_1=5$, $N_2=3$, with non-zero elements $F_{11}=F_{13}=F_{51}=F_{53}$. These are Cartesian products of two one-dimensional Cantor sets with dimensions $D_x=\frac{1}{2}$ ($4\times 3$ case) or $D_x={\rm ln}\,2/{\rm ln}\,5$ ($5\times 3$ case),
and $D_y={\rm ln}\,2/{\rm ln}\,3$. In figure \ref{fig: 4.3} we show the dependence of
$\langle {\cal N}\rangle$ upon $\epsilon$ on a double-logarithmic scale, for the set illustrated
in figure \ref{fig: 2.1}. For each value of $\alpha$, including $\alpha=1$, the plots appear to have an oscillation about a straight line, which makes
it difficult to determine accurate values for $\beta(\alpha)$. In figure \ref{fig: 4.4} we show
our best estimates for $\beta(\alpha)$ for these three fractal sets, compared with the theoretical
prediction in the $4\times 3$ case, given by equation (\ref{eq: 2.2}). There are substantial deviations
from the theoretical curve, but note that these are no larger than the errors in determining the fractal
dimension $D_2=\beta(1)$ for these sets. We conclude that although the agreement with (\ref{eq: 2.2})
is poor, this is due to difficulties in fitting the data, and there is no persuasive evidence that (\ref{eq: 2.2})
is incorrect.

Section \ref{sec: 2} concluded with an argument which suggests that when the
set is locally a Cartesian product of a line and a Cantor set, the function
$\beta(\alpha)$ may have a different behaviour when $\alpha<\frac{1}{2}$
if the lines are curved. The data on the clustering of inertial particles (reported in section \ref{sec: 3}) were not
a good fit to equation (\ref{eq: 2.3}), and they did not show a clear signature
that $\alpha =\frac{1}{2}$ is a critical point.
A more controlled test was made with the Sierpinski fractals considered in this
section. The sets were modified by applying a shift to all of the $y$-coordinates:
\begin{equation}
\label{eq: 4.3}
\mbox{\boldmath$x$}=(x,y)\rightarrow \mbox{\boldmath$x$}'=(x',y')=(x,y+C \sin(2\pi x))
\ .
\end{equation}
The sets which are reported upon in figure \ref{fig: 4.2} were deformed according to this rule,
with $C=\frac{1}{2}$. The $\beta(\alpha)$ exponents remained unchanged for all of these sets
(within numerical fluctuations comparable to those in figure \ref{fig: 4.2}).

\section{Concluding remarks}
\label{sec: 5}

This paper has reported the first systematic study of fractals using a set of covers
which become more anisotropic as they are made smaller. The growth of
the anisotropy is described by a
parameter $\alpha$, and we characterised the
efficiency of covering by a generalised dimension $\beta(\alpha)$. We found that
different sets with the same correlation dimension $D_2$ can have different
$\beta(\alpha)$, as was demonstrated very clearly for the simple Sierpinski
substitution fractals which we considered in section \ref{sec: 4}.

In the case where the fractal is locally a Cartesian product of a line and a Cantor set,
a heuristic argument (similar to that given in \cite{Gra+84} and subsequent papers)
suggest that $\beta(\alpha)$ should  be given by (\ref{eq: 2.3}). We find that
this expression works well for simple model sets of the type considered in section
\ref{sec: 4}, and for the Sinai map, where the attractor is locally a Cartesian product of a line
and a Cantor set.  In the case of clustering of inertial particles, however, we find
that equation (\ref{eq: 2.3}) does not give a good approximation to
$\beta(\alpha)$. In cases where the Cantor set does not have a Cartesian product
structure, we were not able to derive an expression for $\beta(\alpha)$, and we find
persuasive evidence that this function is non-universal.

In three or more dimensions there may also be a tendency for particles to accumulate on
filamentary structures, as well as on planes. This could be characterised by defining
the exponent $\beta$ as a function of two parameters, $\alpha_1$, $\alpha_2$ defining a
covering ellipsoid with principal axes $\epsilon$, $\epsilon^{\alpha_1}$ and $\epsilon^{\alpha_2}$.

This study was partly motivated by the desire to understand the inhomogeneity of light scattering from
fractal distributions of particles, but our investigations indicate that this is a very complex problem, which
will be considered in a separate publication.

{\sl Acknowledgements}. HRK thanks the Open University for a postgraduate
studentship.

\end{document}